\documentstyle[10pt, aas2pp4]{article}
\slugcomment{Submitted to ApJ, June 30, 1997.}
\lefthead{KIM, LEE \& GOODMAN}
\righthead{TWO-COMPONENT GLOBULAR CLUSTERS}
\newcommand\etal{et al. }
\def\spose#1{\hbox to 0pt{#1\hss}}
\newcommand\lsim{\mathrel{\spose{\lower 3.0pt\hbox{$\mathchar"218$}}
     \raise 2.0pt\hbox{$\mathchar"13C$}}}
\newcommand\gsim{\mathrel{\spose{\lower 3.0pt\hbox{$\mathchar"218$}}
     \raise 2.0pt\hbox{$\mathchar"13E$}}}
\newcommand\msun{{\,M_\odot}}
\newcommand\rsun{{\,R_\odot}}

\begin{document}
\twocolumn
\title{TWO-COMPONENT FOKKER-PLANCK MODELS FOR THE EVOLUTION OF
ISOLATED GLOBULAR CLUSTERS}
\author{Sungsoo S. Kim\altaffilmark{1}}
\affil{Institute for Basic Sciences, Pusan National University, Pusan 609-735,
Korea}
\authoremail{sskim@uju.es.pusan.ac.kr}
\author{Hyung Mok Lee}
\affil{Department of Earth Sciences, Pusan National University, Pusan 609-735,
Korea}
\authoremail{hmlee@uju.es.pusan.ac.kr}
\and
\author{Jeremy Goodman}
\affil{Princeton University Observatory, Peyton Hall, Princeton, NJ 08544, USA}
\authoremail{jeremy@astro.princeton.edu}

\altaffiltext{1}{Present address: Dept. of Physics, Space Sci. Lab., Korea
Advanced Institute of Science \& Technology, Daejon 305-701, Korea}

\begin{abstract}
Two-component (normal and degenerate stars) models are the simplest
realization of clusters with a mass spectrum because high mass stars
evolve quickly into degenerates, while low mass stars remain on the
main-sequence for the age of the universe.  Here we examine the
evolution of isolated globular clusters using two-component
Fokker-Planck (FP) models that include heating by binaries formed in
tidal capture and in three-body encounters.  Three-body binary heating
dominates and the postcollapse expansion is self-similar, at least in models
with total mass $M \le 3 \times 10^5 \, M_{\odot}$, initial half-mass radius
$r_{h,i} \ge 5 \, {\rm pc}$, component mass ratio $m_2/m_1 \ge 2$, and
number ratio $N_1/N_2 \le 300$ when $m_2=1.4 \, M_{\odot}$.  We derive
scaling laws for $\rho_c$, $v_c$, $r_c$, and $r_h$ as functions of
$m_1/m_2$, $N$, $M$, and time $t$ from simple energy-balance arguments, and
these agree well with the FP simulations.  We have studied the
conditions under which gravothermal oscillations (GTOs) occur.  If
$E_{tot}$ and $E_c$ are the energies of the cluster and of the core,
respectively, and $t_{rh}$ and $t_c$ are their relaxation times, then
$\epsilon\equiv(E_{tot}/t_{rh})/ (E_c/t_{rc})$ is a good predictor of GTOs:
all models with $\epsilon>0.01$ are stable, and all but one with
$\epsilon<0.01$ oscillate.  We derive a scaling law for $\epsilon$
against $N$ and $m_1/m_2$ and compared with our numerical results.
Clusters with larger $m_2/m_1$ or smaller $N$ are stabler.
\end{abstract}


\section{INTRODUCTION}

The dynamical evolution of pre- and postcollapse globular
clusters is influenced by many factors: the initial mass function,
the nature and efficiency of energy generation mechanisms, galactic tides,
anisotropy of the velocities, the initial population of binaries,
and stellar evolution (see, for example, Spitzer 1987).
If the goal is to model globular clusters realistically, then all of
these effects should be included.
On the other hand, if the goal is a deeper theoretical understanding
of individual dynamical processes, then simpler models can be
more instructive.

A distribution of stellar masses affects the postcollapse
evolution of globular clusters in ways that have not been fully
explored.
The simplest nontrivial multimass models are those with just two
components.
This simplification is drastic but not entirely unrealistic.
Since the stellar main-sequence
lifetime is a very steep function of stellar mass,
a cluster can be assumed to start its dynamical evolution with a 
turnoff-point mass very similar to that currently observed.  Stars
with higher mass have already evolved into degenerate remnants
(such as black holes, neutron stars or white dwarfs).
Because of mass segregation (also called ``mass stratification,'' cf.
Spitzer 1987), the inner parts of a dynamically
relaxed cluster should consist primarily of the turnoff stars and
the heaviest remnants.
The black holes are probably much more massive than present day main-sequence
stars, but their number is expected to be too small
to play any important dynamical role (Kulkarni, Hut, \& McMillan 1993).
Most of the white dwarfs in a cluster are expected to have masses
less than the present turn-off point mass.  
The dynamical effects of such white dwarf stars
are expected to be unimportant.
Kim \& Lee (1997) compared the numerical results of two- and 11-component
models, which have 3 components for the white dwarfs heavier than the turn-off
point mass, and they were able to find two-component cluster parameters that
well match the 11-component clusters.  This implies that the presence of the
white dwarfs heavier than the turn-off point mass does not harm the
two-component approximation.

Neutron stars, however, can be an important component in dynamical evolution
of globular clusters. Neutron stars are about twice as massive as
turnoff stars. The fraction of the cluster mass in the form of
neutron stars is expected to be very small, but mass segregation may
make neutron stars major component in the central parts. (For observational
evidence supporting the dominance of heavy degenerates, see Gebhardt
\& Fischer 1995 and Phinney 1993.)  As a first
approximation, the globular clusters can be represented by two components:
neutron stars and main-sequence stars.

Two-component clusters have several interesting features that are not present
in single-component clusters.  Mass segregation accelerates the early phases
of core collapse significantly (e.g. Yoshizawa \etal 1978; Inagaki \& 
Wiyanto 1984; Spitzer 1987).
The central parts quickly become dominated by 
heavy component, even though bulk of the mass of the cluster is
in the light component.  Tidal captures between a
neutron star and a main-sequence star are frequent, and
binaries composed of neutron stars can also be formed by three-body
processes. These binaries will eventually drive the postcollapse 
evolution.

The purpose of this paper is to obtain a series of solutions for the
dynamical evolution of two-component clusters, with
emphasis on the postcollapse phase.  We study relative
importance of tidal-capture and three-body binary formation for
heating the cluster core.
We are particularly interested in the importance of
unequal stellar masses for gravothermal oscillations.

Our model is quite simple compared to real clusters. Stellar
evolution has not been included; it is most important for the
very early evolution of clusters (Chernoff \& Weinberg 1990,
Drukier 1995).  Primordial binaries could be even more
important than the binaries formed by dynamical processes through the
early postcollapse phases (Gao \etal 1991), but we neglect them.
We have ignored the tidal field of the galaxy, even though tidal limitation
qualitatively alters postcollapse evolution of the cluster mass and radius
(H\'enon 1961, Lee \& Goodman 1995), and tidal shocks may further hasten the
destruction of clusters (Kundic \& Ostriker 1995; Gnedin \& Ostriker 1997).
External tidal fields are probably not important for gravothermal oscillations
and other phenomena pertaining to the core, except insofar as they modify
the total cluster mass.

This paper is organized as follows. In \S \ref{sec:models}, we
describe the models and methods of our calculations. Heating mechanisms
that derive the postcollapse expansion are compared in \S \ref{sec:heating}.
In \S \ref{sec:evolution},
we obtain a series of solutions and give simple analytic expressions for
the evolution of parameters of postcollapse clusters.
Gravothermal oscillation in two-component models is examined in
\S \ref{sec:grav_oscil}. The final section summarizes our results.

\section{MODELS}
\label{sec:models}

\subsection{Computational Method}

The dynamical evolution of collisionless stellar systems under the influence
of two-body relaxation is well described by the Fokker-Planck equation.
We have restricted ourselves to isotropic models, in which the stellar
orbital distribution function depends only on energy (and time).
A recent study by Takahashi, Lee, \& Inagaki (1997) showed that the
radial anisotropy in the halo becomes highly suppressed when a tidal field
is imposed.  The global evolution can be simply described by an
isotropic model.
The multi-component Fokker-Planck equation can be written as follows:
\begin{equation}
	4 \pi^2 p(E) {\partial f_i (E)\over \partial t} = - {\partial F_i (E)
		\over \partial E},
\end{equation}
where $f_i(E)$ and $F_i (E)$ are the distribution function and the particle
flux in energy space $E$, respectively, for the i-th component. 
Formally $F_i$ can be expressed as
\begin{equation}
	F_i (E) = -m_i D_E f_i (E) - D_{EE} {\partial f_i \over \partial E},
\end{equation}
where $D_E$ and $D_{EE}$ are the Fokker-Planck coefficients.  The statistical
weight factor $p(E)$ is given by
\begin{equation}
	p(E)=4\int_0^{\phi^{-1}(E)} \, r^2v \, dr,
\end{equation}
where $\phi^{-1}(E)$ denotes the maximum radius that a particle with energy
$E$ can reach in the spherical potential $\phi(r)$.

In order to account for the effects of binaries, we need to modify the
Fokker-Planck equation above. Statler, Ostriker \& Cohn (1987) developed 
a sophisticated  scheme to include the dynamical effects of tidal
binaries for a model initially composed of a single component.
Lee (1987) modified the method of Statler \etal (1987) to include both
tidal binaries and three-body binaries. However, the number of
dynamically-produced binaries
is always a very small fraction of the total number of stars. The most important
effect is heating---the addition of entropy to the orbital distribution
function---which can be simply accounted for by modifying the particle
flux in energy space. Such an approach has been taken in many studies 
dealing with the three-body binary heating (see, for example, Cohn 1985;
Lee, Fahlman, \& Richer 1991). Normally the tidal
binaries are more abundant and simple correction of the particle flux could
cause some errors. However, we find that our approach provide excellent 
agreement with more complex schemes.

The orbit-averaged heating coefficient due to this heating becomes
\begin{equation}
	H_i (E) = { \int _0^{r_{\rm max}} \dot E vr^2 \, dr \over
		  \int_0^{r_{\rm max}} vr^2 \, dr },
\end{equation}
where $\dot E$ is the heating rate per unit volume and $r_{\rm max}$ the
maximum radius accessible to a star with energy $E$.
The modified particle flux then takes the form
\begin{equation}
	F_i (E) = -\left[m_i D_E + H_i (E)\right] f_i (E) -D_{EE} {\partial f_i \over
		  \partial E}.
\end{equation}
The second-order coefficient ($D_{EE}$) is also affected
by binary heating, but this diffusive effect
is probably much less important for the long-term evolution than the
change in $D_E$.

Heating by tidal binaries is mostly due to their ejection during
close encounters with single stars (or sometimes other binaries)
because the internal
binding energy of tidal binaries is much greater than the escape 
energy from the cluster. Therefore the heating rate per unit volume
is the product of the mass in all three stars, the binary formation rate,
and the central potential (Lee \& Ostriker 1993):
\begin{equation}
\label{Edottc}
	{\dot E}_{tc}=(m_n+2m_d) \sigma_{tc} v_{rel} n_n n_d \phi_c,
\end{equation}
where $n$ is the number density, $\sigma_{tc}$ the tidal-capture cross
section, $v_{rel}$ the relative rms velocity between degenerate and normal
stars, and $\phi_c$ the central gravitational potential.  The subscript $n$
denotes normal stars while $d$ represents degenerate stars.
It is assumed that the binary is ejected promptly after formation by
interaction with a third star.
We adopt the following expression for $\sigma_{tc}$ from Lee \& Ostriker
(1986) for tidal captures between a normal star and a degenerate star with
$m_d/m_n=2$:
\begin{equation}
	\sigma_{tc} = 13 \left ( {v_{rel} \over v_{e,n}} \right )^{-2.1}
		      R_n^2,
\label{eq:sigtc}
\end{equation}
where $v_{e,n}$ is the escape velocity at the normal star's surface, and
$R_n$ the normal star's radius.

Binaries formed by dissipationless three-body encounters (``three-body
binaries'') are usually less tightly bound than tidal binaries.
Therefore, in their encounters with single stars, some or all of the
stars are retained by the cluster and contribute their increased
orbital energy to the distribution functions
$f_i$ (``direct heating'').
Until ejected from the cluster, a three-body binary releases approximately
300 $kT$, where
$kT$ is the typical kinetic energy of background stars. Thus the three-body
binary heating rate per unit volume can be computed by multiplying the binary
formation rate per unit volume by 300 $kT$:
\begin{equation}
\label{Edot3b}
	{\dot E}_{3b} = 4.21 \times 10^3 \, G^5 \left ( \sum_i
			{n_i m_i^2 \over v_i^3} \right )^3 v_c^2, 
\end{equation}
where the summation is over all components, and $v_c^2$ is the mass weighted,
three-dimensional, central velocity dispersion.  The numerical
coefficient has been taken from Cohn (1985).

The modified Fokker-Planck equation can be accurately integrated with the
numerical procedure described by Cohn (1979, 1980).

\subsection{Model Parameters}

Our models assume an initial cluster composed of main-sequence stars with mass
$m_1$ and neutron stars with mass $m_2$. The total masses in the form
of these stars are $M_1$ and $M_2$ respectively. The total number of stars
is denoted by $N$.

We need to specify three dimensionless parameters in addition to the 
initial density and velocity profiles:  the total number of stars $N$, the
ratio $m_2/m_1$ of a heavy star to a light one, and the number ratio $N_2/N_1$.
The dimensional scales are determined by the total cluster mass $M$, 
the initial half-mass radius $r_{h,i}$, and the
stellar radius $R_n$.  The ratio $m_1 r_{h,i}/M R_n$ is important for
tidal heating because it determines the ratio of the capture cross
section (eq. [\ref{eq:sigtc}]) to the projected area of the cluster 
(or cluster core).  The present study may be devided into three
topics and each topic has its own set of models.  The detailed model
parameters of those sets are listed in Tables \ref{table:heating},
\ref{table:A}, and \ref{table:B}.
Note that in all our runs, the total mass of the heavy component, 
$M_2\equiv N_2 m_2$, is negligible compared to the total mass of light
component, $M_1\equiv N_1 m_1$, and thus
$m_1 \approx M/N_1$.  The initial density and velocity profiles are given
by Plummer models with $v_{c1}/v_{c2}=1$ and $\rho_{c1}/\rho_{c2}=M_1/M_2$,
where $\rho_c$ is the core density.  Both three-body binary heating and
tidal binary heating are included and clusters are assumed to be
isolated (i.e. no tidal cut-off).

Our values for $m_2/m_1$ range from 2 to 5. If we identify
$m_1$ with the heaviest main-sequence stars, then $m_2/m_1 = 2$ is a
suitable choice. Since the bulk of the cluster consists of stars below
the turnoff, however,  $m_1$ should be somewhat smaller than the
turnoff-point mass.
Finally, we  assume a linear mass-radius relation such that $1\,\msun$
corresponds to $1\,\rsun$.

\section{HEATING MECHANISMS}
\label{sec:heating}

Most heating takes place in the core where the stellar and binary
number densities are highest.  
To compare the efficiencies of heating mechanisms, we first derive
their dependence on core parameters.
From equations (\ref{Edottc}) and (\ref{Edot3b}), $\dot E_{tc}$
in the core is approximately proportional to $v_n n_n n_d$, while
$\dot E_{3b}$ in a core dominated by degenerate stars is approximately
proportional to $n_d^3 v_d^{-7}$.  We have assumed
that $\phi_c \propto v_{n,c}^2$, which is valid during the
postcollapse phase.  Even during the precollapse phase, the ratio of
potential depth to central velocity dispersion varies very slowly compared
with other quantities.  Considering again the fact that the core is dominated
by degenerate stars, the central heating rate by tidal binaries is
proportional to the first power of the central density, and
the central heating rate by three-body binaries is proportional to the third
power. Since the central velocity dispersion evolves much less than central
density, the density of degenerate stars in the core is the most important
quantity in deciding the relative efficiency of the two heating
mechanisms. Thus the relative importance of three-body binaries 
increases as the core collapse proceeds.

\begin{deluxetable}{rccccccrcccc}
\footnotesize
\tablecolumns{12}
\tablewidth{0pt}
\tablecaption{Dominant Heating Mechanism in the Postcollapse Phase
\label{table:heating}}
\tablehead{
\colhead{$M$} &
\multicolumn{4}{c}{$r_{h,i}$ (pc)} &
\colhead{} &
\colhead{} &
\colhead{$M$} &
\multicolumn{4}{c}{$r_{h,i}$ (pc)} \\
\cline{2-5} \cline{9-12}
\colhead{($\msun$)} &
\colhead{$1$} &
\colhead{$2.5$} &
\colhead{$5$} &
\colhead{$10$} &
\colhead{} &
\colhead{} &
\colhead{($\msun$)} &
\colhead{$1$} &
\colhead{$2.5$} &
\colhead{$5$} &
\colhead{$10$}
}
\startdata
\tablevspace{0.5ex}
&\multicolumn{4}{c}{${m_2 \over m_1}=2$, ${N_1 \over N_2}=30$} &&&
&\multicolumn{4}{c}{${m_2 \over m_1}=2$, ${N_1 \over N_2}=300$} \nl
\tablevspace{0.5ex}
\cline{2-5} \cline{9-12}
$1 \times 10^5$ & 3B & 3B & 3B & 3B &&& $1 \times 10^5$ & TB & 3B & 3B & 3B \nl
$3 \times 10^5$ & 3B & 3B & 3B & 3B &&& $3 \times 10^5$ & TB & TB & 3B & 3B \nl
$1 \times 10^6$ & TB & 3B & 3B & 3B &&& $1 \times 10^6$ & TB & TB & TB & TB \nl
$3 \times 10^6$ & TB & TB & 3B & 3B &&& $3 \times 10^6$ & TB & TB & TB & TB \nl
\tablevspace{1ex}
&\multicolumn{4}{c}{${m_2 \over m_1}=3$, ${N_1 \over N_2}=30$} &&&
&\multicolumn{4}{c}{${m_2 \over m_1}=3$, ${N_1 \over N_2}=300$} \nl
\tablevspace{0.5ex}
\cline{2-5} \cline{9-12}
$1 \times 10^5$ & 3B & 3B & 3B & 3B &&& $1 \times 10^5$ & 3B & 3B & 3B & 3B \nl
$3 \times 10^5$ & 3B & 3B & 3B & 3B &&& $3 \times 10^5$ & 3B & 3B & 3B & 3B \nl
$1 \times 10^6$ & 3B & 3B & 3B & 3B &&& $1 \times 10^6$ & TB & 3B & 3B & 3B \nl
$3 \times 10^6$ & 3B & 3B & 3B & 3B &&& $3 \times 10^6$ & TB & TB & TB & 3B \nl
\tablecomments{3B is for the models whose postcollapse expansion is driven
by three-body binary heating, and TB by tidal binary heating.  $m_2 = 1.4 \msun$
for all models.}
\enddata
\end{deluxetable}

\begin{figure}[t]
\plotone{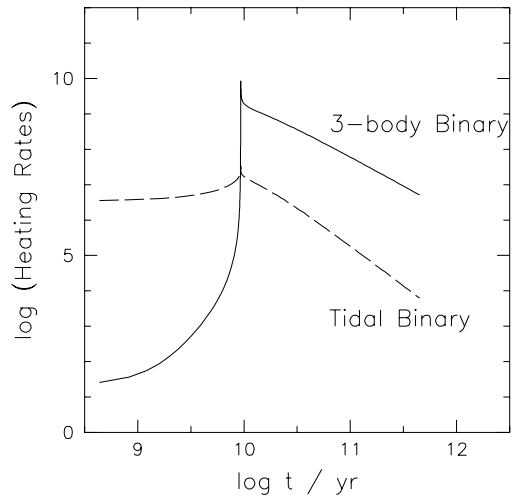}
\caption
{\label{fig:heating}Heating rates by three-body binaries (solid line)
and tidal binaries (dashed line) of run caab.  The units of the y-axis are
arbitrarily chosen.}
\end{figure}

We have searched for the division of tidal and three-body heating dominance
in the two-component parameter space, and its result is shown in Table
\ref{table:heating}.  3B denotes the models whose postcollapse expansion is
driven by three-body binaries, and TB by tidal binaries.
Three-body binary heating is relatively more important for clusters with
smaller $M$ and $N_1/N_2$, and larger $r_{h,i}$ and $m_2/m_1$.  Smaller
$N_1/N_2$ and larger $m_2/m_1$ give larger initial $n_{c2}/n_{c1}$ which is
favorable to three-body binary formation.  On the other hand, a cluster with
very high initial $n_c$ starts its evolution with tidal binary heating rate
high enough not to give a change for three-body binary heating to be dominant
during the collapse.  Kim \& Lee (1997) found that the evolution of
multi-component models with various power-law mass functions and $M$ may be
realized with two-component models with $1.7 < m_2 < 3$ and $10 < N_1/N_2 < 50$
(the actual fraction of neutron stars in globular clusters is much less than
1/100, but many low-mass normal stars do not contribute much on the cluster's
dynamical evolution).  Although the dominant heating
mechanism is dependent on the mass function ($m_2/m_1$ and $N_1/N_2$),
all models with typical globular cluster masses and sizes
($M \lsim 10^6 \msun$ and $r_{h,i} \gsim 2.5 \, {\rm pc}$) and with
the above mass function range are dominated by three-body binary heating
in the postcollapse phase as in Figure \ref{fig:heating}, which is a typical
history of the heating rates of our models marked with 3B in Table
\ref{table:heating}.  This dominance of 3B over TB in most plausible
parameter range confirms the conclusion of Lee (1987) that in
multi-component clusters, three-body binary heating is relatively more
important because there are no tidal captures between pairs of degenerate
stars.  Note that, even for an extreme $N_1/N_2$ value of 300, three-body
binary heating is still dominant for clusters with $M \le 3 \times 10^5 \msun$
and $r_{h,i} \ge 5 \, {\rm pc}$, which are the parameter ranges that the
majority of the present clusters have.

\begin{figure}[t]
\plotone{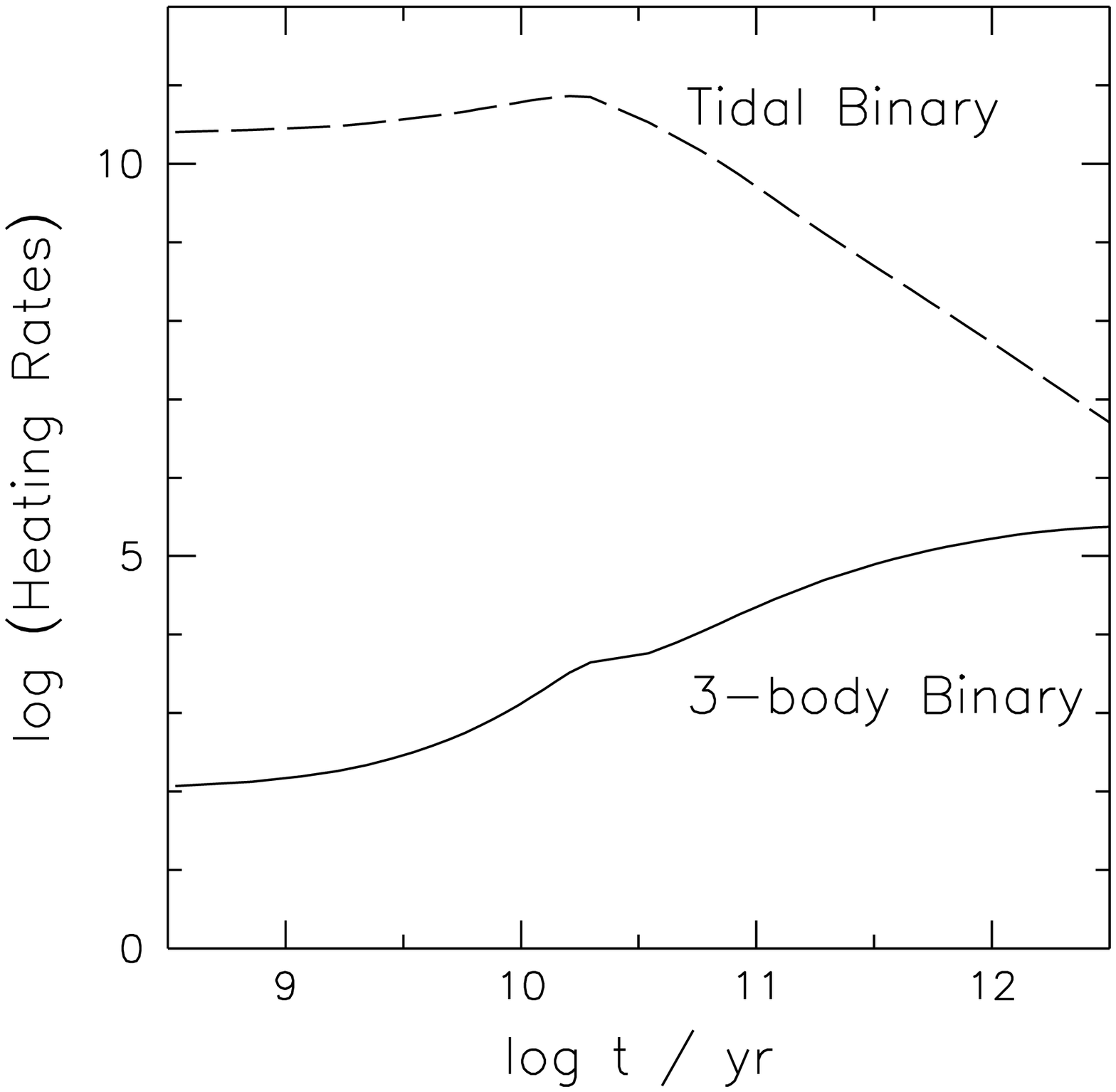}
\caption
{\label{fig:heating_tc}Heating rates by three-body binaries (solid line)
and tidal binaries (dashed line) of our two-component model with
$M=10^6 \msun$, $r_{h,i}=2.5{\rm pc}$, $m_2/m_1=2$, $N_1/N_2=300$, and
$m_2=1.4 \msun$.  The units of the y-axis are arbitrarily chosen.  The initial
tidal binary heating rate is so high that the three-body binary heating does
not have a chance to take over.  However, the three-body binary
heating keeps increasing in the postcollapse phase.}
\end{figure}

\begin{figure}[t]
\plotone{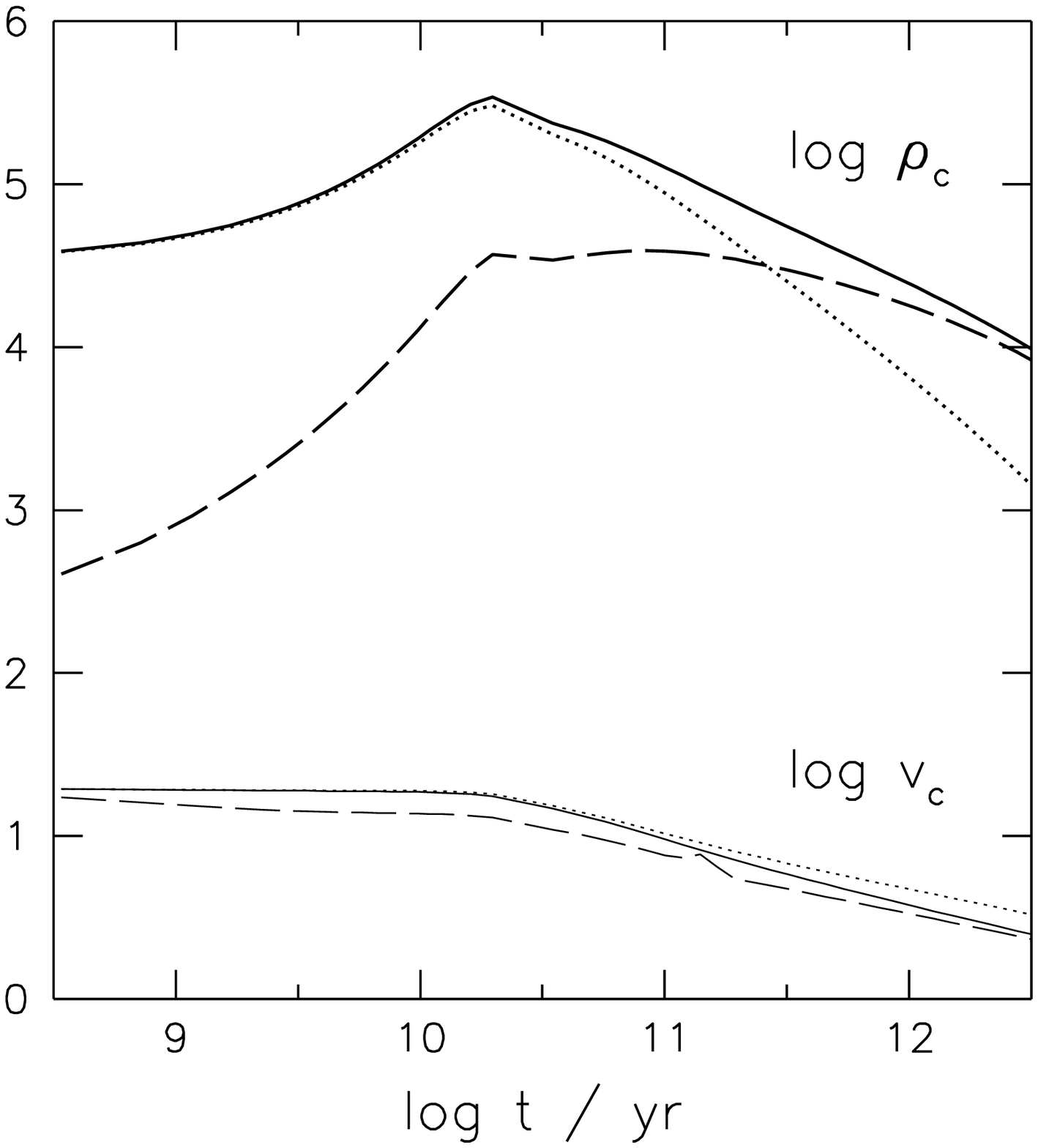}
\caption
{\label{fig:rhov_tc}Temporal evolution of the core density (thick lines)
and the central velocity dispersion (thin lines) for the same run as in Figure
\protect\ref{fig:heating_tc}.  Dotted lines are for the light component, dashed lines
for the heavy component, and solid lines for the overall values.  Equipartition
is approached in the postcollapse phase.  $\rho_c$ is in units of $M_\odot
{\rm pc^{-3}}$ and $v_c$ in ${\rm km\, s^{-1}}$.}
\end{figure}

Although tidal binary heating is unimportant for postcollapse
expansion of models marked with 3B, 
tidal binary heating becomes more important when $M$ and $N_1/N_2$
are larger, and when $r_{h,i}$ and $m_2/m_1$ are smaller.  For example, our
two-component model with $M=10^6 \msun$, $r_{h,i}=2.5\,{\rm pc}$,
$m_2/m_1=2$, $N_1/N_2=300$, and $m_2=1.4 \msun$ is governed
by tidal binary heating in the postcollapse phase (see Figure
\ref{fig:heating_tc}).  It is notable, however, that the three-body
binary heating rate increases after core collapse while the rate of
heating by tidal binaries is decreasing. This phenomenon is
not expected for single-mass
tidal-binary-driven postcollapse clusters, in which the central density
and velocity dispersion evolve as $\rho_c \propto t^{-1.04}$ and
$v_c \propto t^{-0.34}$ (Lee \& Ostriker 1993), and the three-body
binary heating rate as $\rho_c^3 v_c^{-7} \propto t^{-0.8}$.
In a two-component postcollapse cluster, however, if tidal binaries
dominate and equipartition holds
in the core, then the central density decreases less rapidly and the
central velocity more rapidly than in a single-component cluster
(Figure \ref{fig:rhov_tc}).  In such cases, the heating rate by
three-body binaries may even increase during postcollapse expansion.

The ultimate age of the models shown in Figures
\ref{fig:heating}--\ref{fig:converge} appears unrealistically long.
But the evolution of isolated clusters
slows down as $r_h$ expands (see \S 3.3 below),
whereas $r_h$, $t_{rh}$, and $M$ decrease during the postcollapse evolution
of tidally limited clusters, which are more realistic
(cf. Lee \& Ostriker 1987).
Thus, the evolutionary state of an isolated cluster at $t=10^{11}~{\rm yr}$
corresponds roughly to that of a tidally limited one at a much earlier
time.  Mass loss by overflow of the tidal boundary is much 
more rapid than ejection of stars by binaries in the core, and depletes
primarily the lighter component.  By decreasing $N_1/N_2$, this last
effect further suppresses tidal binaries.

\begin{deluxetable}{rccccccrcccc}
\footnotesize
\tablecolumns{12}
\tablewidth{0pt}
\tablecaption{Dominant Heating Mechanism in the Postcollapse Phase (Merging
Case) \label{table:heating_merge}}
\tablehead{
\colhead{$M$} &
\multicolumn{4}{c}{$r_{h,i}$ (pc)} &
\colhead{} &
\colhead{} &
\colhead{$M$} &
\multicolumn{4}{c}{$r_{h,i}$ (pc)} \\
\cline{2-5} \cline{9-12}
\colhead{($\msun$)} &
\colhead{$1$} &
\colhead{$2.5$} &
\colhead{$5$} &
\colhead{$10$} &
\colhead{} &
\colhead{} &
\colhead{($\msun$)} &
\colhead{$1$} &
\colhead{$2.5$} &
\colhead{$5$} &
\colhead{$10$}
}
\startdata
\tablevspace{0.5ex}
&\multicolumn{4}{c}{${m_2 \over m_1}=2$, ${N_1 \over N_2}=30$} &&&
&\multicolumn{4}{c}{${m_2 \over m_1}=2$, ${N_1 \over N_2}=300$} \nl
\tablevspace{0.5ex}
\cline{2-5} \cline{9-12}
$1 \times 10^5$ & 3B & 3B & 3B & 3B &&& $1 \times 10^5$ & 3B & 3B & 3B & 3B \nl
$3 \times 10^5$ & 3B & 3B & 3B & 3B &&& $3 \times 10^5$ & TB & 3B & 3B & 3B \nl
$1 \times 10^6$ & 3B & 3B & 3B & 3B &&& $1 \times 10^6$ & TB & TB & 3B & 3B \nl
$3 \times 10^6$ & 3B & 3B & 3B & 3B &&& $3 \times 10^6$ & TB & TB & TB & TB \nl
\tablevspace{1ex}
&\multicolumn{4}{c}{${m_2 \over m_1}=3$, ${N_1 \over N_2}=30$} &&&
&\multicolumn{4}{c}{${m_2 \over m_1}=3$, ${N_1 \over N_2}=300$} \nl
\tablevspace{0.5ex}
\cline{2-5} \cline{9-12}
$1 \times 10^5$ & 3B & 3B & 3B & 3B &&& $1 \times 10^5$ & 3B & 3B & 3B & 3B \nl
$3 \times 10^5$ & 3B & 3B & 3B & 3B &&& $3 \times 10^5$ & 3B & 3B & 3B & 3B \nl
$1 \times 10^6$ & 3B & 3B & 3B & 3B &&& $1 \times 10^6$ & 3B & 3B & 3B & 3B \nl
$3 \times 10^6$ & 3B & 3B & 3B & 3B &&& $3 \times 10^6$ & TB & 3B & 3B & 3B \nl
\tablecomments{Notations and $m_2$ are the same as in Table \protect\ref{table:heating}.
}
\enddata
\end{deluxetable}

Many of the interactions that we identify as tidal captures would
actually have lead to mergers (Benz \& Hills 1987).
Ultimately, the mass of the normal star might still be ejected,
but the neutron star would probably not be.
Thus the parameter space for clusters whose postcollapse phase is driven by
tidal binary heating is expected to be even narrower than our present
calculations suggest.
As a limiting case, we have re-calculated the two-component models in Table
\ref{table:heating} assuming that the all tidal capture binaries end up with
mergers and only the masses of the normal stars are ejected, i.e.
\begin{equation}
\label{Edottc2}
        {\dot E}_{tc}= m_n \sigma_{tc} v_{rel} n_n n_d \phi_c.
\end{equation}
The dominant heating mechanism in the postcollapse phase for the models in
Table \ref{table:heating} with the above ${\dot E}_{tc}$ is shown in Table
\ref{table:heating_merge}.  The dominance of tidal binary heating is seen
only for few models, because equation (\ref{Edottc2}) gives $1/(2m_2/m_1 +1)$
times less heating rate than equation (\ref{Edottc}) and thus three-body
binaries now become even more important.  For realistic mass functions
($N_1/N_2 \le 50$), no model in our parameter range ($M \le 3 \times 10^6$ and
$r_{h,i}> \, {\rm pc}$) is marked with TB.

\section{PRE- AND POSTCOLLAPSE EVOLUTION}
\label{sec:evolution}

In this section, we show the results for 11 models (Group A).
The parameters of these models, which are shown in Table
\ref{table:A}, have been chosen to cover a range of plausible or instructive
combinations.  We now discuss some features of these models.

\begin{deluxetable}{lrrrrrrcclc}
\tablecolumns{11}
\tablewidth{0pt}
\tablecaption{Parameters and Results of Simulation Group A \label{table:A}}
\tablehead{
\multicolumn{7}{c}{} &
\multicolumn{4}{c}{Values at $t=10^{11}$yr} \\
\cline{8-11}
\colhead{Run} &
\colhead{$m_2 \over m_1$} &
\colhead{$N_1 \over N_2$} &
\colhead{$M$} &
\colhead{$N$} &
\colhead{$m_2$} &
\colhead{$t_{cc} / t_{rh,i}$} &
\colhead{$\rho_c$} &
\colhead{$v_c$} &
\colhead{$r_c$} &
\colhead{$r_h$} \\
\multicolumn{3}{c}{} &
\colhead{($\msun$)} &
\colhead{} &
\colhead{($\msun$)} &
\colhead{} &
\colhead{($\msun \, {\rm pc^{-3}}$)} &
\colhead{(${\rm km\,s^{-1}}$)} &
\colhead{(pc)} &
\colhead{(pc)}
}
\startdata
baab  &2 &100 &       $10^5$ &141457 &                 1.4 &12.42&$1.47\times 10
^5$ &3.03 &0.0579 &28.8\nl
caab  &3 &100 &       $10^5$ &210125 &                 1.4 & 6.34&$1.29\times 10
^5$ &2.86 &0.0585 &21.2\nl
faab  &4 &100 &       $10^5$ &277473 &                 1.4 & 3.23&$1.19\times 10
^5$ &2.80 &0.0594 &19.9\nl
cdab  &3 & 30 &       $10^5$ &201299 &                 1.4 & 3.97&$1.26\times 10
^5$ &2.90 &0.0600 &28.6\nl
cbab  &3 &300 &       $10^5$ &212871 &                 1.4 &10.92&$1.26\times 10
^5$ &2.85 &0.0587 &21.5\nl
caab1 &3 &100 &       $10^5$ & 70042 &        $3\times$1.4 & 6.47&$3.18\times 10
^3$ &2.08 &0.270  &40.9\nl
caab2 &3 &100 &       $10^5$ &630374 &${1\over3}\times$1.4 & 6.28&$5.36\times 10
^6$ &3.91 &0.0124 &11.4\nl
baab3 &2 &100 &       $10^5$ &212185 &${2\over3}\times$1.4 &12.17&$6.10\times 10
^5$ &3.40 &0.0319 &22.7\nl
faab3 &4 &100 &       $10^5$ &208104 &${4\over3}\times$1.4 & 3.23&$4.57\times 10
^2$ &2.58 &0.0884 &23.5\nl
caeb  &3 &100 &$3\times10^4$ & 63037 &                 1.4 & 6.34&$2.21\times 10
^3$ &1.35 &0.210  &29.5\nl
cabb  &3 &100 &$3\times10^5$ &630374 &                 1.4 & 6.43&$5.61\times 10
^6$ &5.68 &0.0176 &15.9\nl
\tablecomments{The initial half-mass radii $r_{h,i}$ of these runs
are all 5 pc.}
\enddata
\end{deluxetable}

\subsection{Epoch of Corecollapse}

Isolated, single-component clusters beginning as Plummer models reach
core collapse after a time
$t_{cc}=15.4~t_{rh,i}$ (Cohn 1980), where $t_{rh,i}$, the initial half-mass
relaxation time, does not vary much before core collapse.
However, the ratios $t_{cc}/t_{rh,i}$ and $t_{cc}/t_{rc}$ (where
$t_{rc}$ is the core relaxation time) depend strongly on the
initial density and velocity profile (Inagaki 1985), and can be
much smaller for choices other than the conventional Plummer model.
Quinlan (1996) found that for single-mass
clusters, $t_{cc}$ varies much less when expressed in units of $t_{rc}$
divided by a dimensionless measure of the temperature gradient in the core.
Single-mass clusters evolve by radial transport of energy, but energy
exchange between different mass components plays an important role in
multi-component models. Mass segregation in a two-component model takes
place initially on a timescale set by dynamical friction, which
can be shorter than the corecollapse time 
of a single-component models with similar macroscopic properties.
Mass segregation and core collapse in two-component models has
been discussed in detail by Inagaki (1985).
In Table \ref{table:A}, we have listed the $t_{cc}/t_{rh,i}$ ratios for each of
our models.

\subsection{Scaling Laws for Postcollapse Evolution}

The expansion of the core in postcollapse is determined by
the dominant heating mechanism.
Here we present scaling laws for the postcollapse evolution of two-component
clusters driven by three-body binary heating based on theoretical analysis
and compare them with our numerical results (see Lee \& Ostriker 1993 for
scaling laws for evolution driven by tidal binary heating).

We start with Goodman's (1993) energy balance analysis arguments.
The energy generation rate by three-body binaries in the core is
\begin{equation}
\label{Lc}
	L_c \approx M_c { \dot E_{3b,c} \over \rho_c},
\end{equation}
where the core mass $M_c$ and core radius $r_c$ are given by
\begin{equation}
\label{Mc}
	M_c \equiv {2 \pi \over 3} \rho_c r_c^3
\end{equation}
\begin{equation}
\label{rc}
	r_c^2 \equiv {3 v_c^2 \over 4 \pi G \rho_c}.
\end{equation}
On the other hand, the power required by the expansion of the cluster is
\begin{equation}
\label{Edoth}
	\dot E_h \approx {G M^2 \over r_h^2} \dot r_h.
\end{equation}
Since the heavy component dominates in the core and the light component at $r_h$,
substitution of the core parameters into equation (\ref{Edot3b}) yields
\begin{equation}
\label{Lc2}
	L_c \propto m_2^3 r_c^3 \rho_c^3 v_c^{-7}
\end{equation}

Slowly evolving postcollapse solutions are almost isothermal and the
core approaches equipartition.  Thus the Virial relation becomes
\begin{equation}
\label{vc}
	v_c^2 \approx v_{c2}^2 \approx {m_1 \over m_2} v_{c1}^2
	      \approx {m_1 \over m_2} {GM \over r_h}.
\end{equation}
Furthermore, the temporal dependence of the above parameters can be obtained
from the assumption that
\begin{equation}
\label{rh-t}
	{\dot r_h \over r_h} \propto {1 \over t_{rh}},
\end{equation}
which would follow if the evolution were self-similar.
Since $t_{rh}\propto M^{1/2} m_1^{-1} r_h^{3/2}$, the above relation gives
\begin{equation}
\label{rhcc-tcc}
	(r_h^{3/2}-r_{h,cc}^{3/2}) \propto {m_1 \over M^{1/2}} (t-t_{cc}),
\end{equation}
where $r_{h,cc}$ is $r_h$ at $t=t_{cc}$, and the outer part of the cluster is
assumed to be dominated by the light component.  Since $r_h$ is almost constant
until the corecollapse takes place,
\begin{equation}
\label{rhcc-tcc2}
	(r_h^{3/2}-r_{h,i}^{3/2}) \propto {m_1 \over M^{1/2}} (t-t_{cc}),
\end{equation}
and for $t \gg t_{cc}$
\begin{equation}
\label{rh-t2}
	r_h^{3/2} \propto {m_1 \over M^{1/2}} t.
\end{equation}

It follows from equations (\ref{rh-t}) and (\ref{Edoth}) that
\begin{equation}
\label{Edoth2}
	\dot E_h \propto m_1 M^{3/2} r_h^{-5/2}.
\end{equation}
Demanding that the power (\ref{Edoth2}) required by expansion be supplied
by the core luminosity (\ref{Lc2}), and substituting
for $r_h$ from (\ref{rh-t2}),
one finds the following relations for late postcollapse
evolution (i.e. $t \gg t_{cc}$):
\begin{mathletters}
\label{scale}
\footnotesize
\begin{eqnarray}
        \rho_c & \propto & \left({m_2 \over m_1} \right)^{-10/3}
                N^{10/3} t^{-2};\\
        v_c   & \propto & \left({m_2 \over m_1} \right)^{-1/2}
                N^{1/3} M^{1/3} t^{-1/3};\\
        r_c   & \propto & \left({m_2 \over m_1} \right)^{7/6}
                N^{-4/3} M^{1/3} t^{2/3};\\
        r_h   & \propto & N^{-2/3} M^{1/3} t^{2/3};\\
	M_c   & \propto & \left( {m_2\over m_1}\right)^{1/6} N^{-2/3} M.
\end{eqnarray}
\end{mathletters}
It is notable that $r_{h,i}$, one of the initial parameters of the cluster,
is not included anywhere in equation (\ref{scale}).  This, too, is a
reflection of the self-similar nature of postcollapse evolution, which has been
confirmed by many previous studies, but mostly for a single mass component.
Tidally limited postcollapse evolution, on the other hand, is not strictly
self-similar, since $r_c/r_h$ increases with decreasing cluster mass.

Unlike tidal binary heating, the calculation of three-body binary heating
can be done purely dimensionlessly and is scalable to isolated clusters
of any initial size, mass, and density.
We can fix all of the constants in the scalings above
from our numerical experiments:
\begin{mathletters}
\label{scale_sim}
\footnotesize
\begin{eqnarray}
	\rho_c & \simeq & 4.5 \times 10^5 \, \msun / {\rm pc^3} \, \,
		\left ({m_2 \over m_1} \right )^{-10/3}
		N_5^{10/3} t_{11}^{-2.0}; \\
	v_c   & \simeq & 3.8 \, {\rm km/s} \, \,
		\left ({m_2 \over m_1} \right )^{-1/2}
		N_5^{1/3} M_5^{1/3} t_{11}^{-0.32}; \\
	r_c   & \simeq & 0.042 \, {\rm pc} \, \,
		\left ({m_2 \over m_1} \right )^{7/6}
		N_5^{-4/3} M_5^{1/3} t_{11}^{0.65}; \\
	r_h   & \simeq & 35 \, {\rm pc} \, \,
		N_5^{-2/3} M_5^{1/3} t_{11}^{0.65}; \\
	M_c   & \simeq & 71\msun \left({m_2\over m_1}
                \right)^{1/6} N_5^{-2/3} M_5 t_{11}^{-0.05},
	\label{eq:mcore}
\end{eqnarray}
\end{mathletters}
where $N_5 \equiv N/10^5$, $M_5 \equiv M/10^5\msun$, and $t_{11} \equiv t/
10^{11}\,{\rm yr}$.
Note that the exponents of $t$, which are obtained by the power-law fitting,
show only small discrepancies from equation (\ref{scale}).

\begin{figure}[t]
\plotone{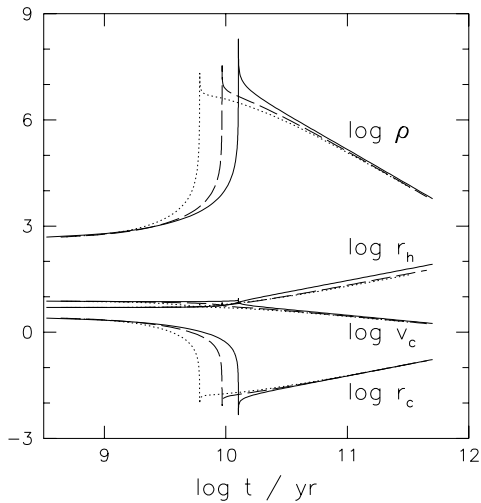}
\caption
{\label{fig:converge}Temporal evolution of
central densities, three-dimensional rms central velocities,
core radii and half-mass radii for runs baab (solid line), caab (dashed line),
and faab (dotted line).  These three runs have nearly the same
$N \times m_1/m_2$ values and converge into the same evolutionary track in
the postcollapse phase.  $\rho_c$ is in units of $M_\odot {\rm pc^{-3}}$,
$r_h$ and $r_c$ in pc, and $v_c$ in ${\rm km\, s^{-1}}$.}
\end{figure}

The postcollapse central density evolutions of some of our runs with the
same initial conditions and the same $N \times m_1/m_2$ values
are shown in Figure \ref{fig:converge}.  According to equation (\ref{scale}),
these runs should have the same $\rho_c$ at the same $t(>t_{cc})$ although
their $t_{cc}$'s are different.  After a short transition period in the
beginning of the postcollapse phase, the $\rho_c$'s do indeed converge to
the same log-log slope.

\begin{figure}[t]
\plotone{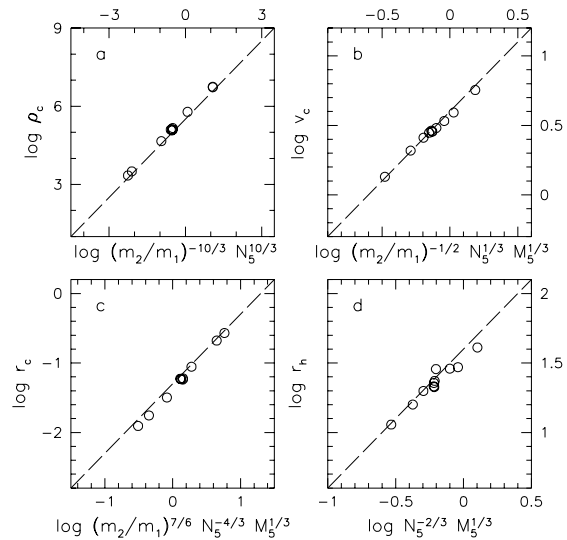}
\caption
{\label{fig:scale}(a) Central densities, (b) three-dimensional rms
central velocities, (c) core radii, and (d) half-mass radii of runs in Group A.
$N \equiv N/10^5$ and $M_5 \equiv M/10^5 \msun$. The straight diagonal lines
represent the theoretical relations.}
\end{figure}

Figure \ref{fig:scale} shows $\rho_c$, $v_c$, $r_c$, and $r_h$ of our runs
in Group A at $t=10^{11}$ yr over the right-hand-sides in equation
(\ref{scale}).  All panels in the figure have the same abscissa and
ordinate scales so that a slope of unity
represents the proportions in equation (\ref{scale}).  As one can see from
these figures, our numerical results closely follow the scaling
relationship. The best-fitting slopes in the
figures are $1.02\pm0.01$, $0.92\pm0.01$, $1.07\pm0.01$, and $0.86\pm0.06$ for
$\rho_c$, $v_c$, $r_c$, and $r_h$, respectively.  These slopes are close to
unity as represented by diagonal lines.
The slope of $r_h$ is the most discrepant.  This is 
probably due to the approximations used in equations (\ref{rh-t}) through
(\ref{rh-t2}).
The validity of equation (\ref{rh-t})
will be discussed again in \S \ref{sec:grav_oscil}.

According to the scalings above, the central density depends
upon the total mass $M$, the epoch $t$, and on $m_2$, but
it is independent of $M_1$ and $M_2$ separately.
Therefore, the mass function does not influence the central density
in postcollapse (provided that there are enough heavy stars to fill
the core---see below), even though it strongly influences
the epoch of core collapse.

The degree of concentration can be measured by $r_h/r_c$ which is
constant during the postcollapse phase. From equations (27c) and 
(27d) we have
\begin{equation}
r_h/r_c = 810 \, N_5^{2/3} \left({m_1\over m_2}\right)^{7/6}.
\end{equation}
In tidally limited postcollapse models, the tidal radius is larger than 
$r_h$ by a factor of the order
of 10. This means the concentration parameter of postcollapse
cluster could become as large as 4. This is consistent
with the fact that stellar density profile of M15 obtained
by HST does not show flat core down to 2$^{\prime\prime}$ (Guhathakurta 
et al. 1996). 
Note, however, that the core radius obtained from the light profile is 
somewhat larger than the core radius of the mass; more importantly, real
clusters are tidally limited and therefore are not described accurately by
these models.

\section{GRAVOTHERMAL OSCILLATIONS}
\label{sec:grav_oscil}

The inner regions of a slowly evolving postcollapse cluster are nearly
isothermal.
Their slow expansion on the timescale $t_{rh}$ is almost negligible
on the local timescale $t_{rc}$.
Thus these regions resemble  equilibrium isothermal spheres, and therefore
they are subject to the gravothermal instability studied by
Antonov (1962) and Lynden-Bell \& Wood (1968).
This instability of postcollapse clusters was first found
by Sugimoto \& Bettwieser (1983) and Bettwieser \& Sugimoto
(1984), and later investigated by a number of scientists.

\subsection{The Instability Parameter $\epsilon$}

Single-component, isolated clusters in postcollapse form a one-parameter family
when they are dominated by three-body binaries.  That
parameter can be taken to be $N$, the number of stars.
In this case, gravothermal instability occurs for
$N \la N_{crit} \approx 7000$ (Goodman 1987).
The introduction of multiple components brings additional dimensionless
parameters, which affect the energy-generation rate in the core and
perhaps also the relaxation rate near $r_h$.
We may therefore expect that the instability should depend on 
parameters other than $N$.

Goodman (1993) suggested that the quantity
\begin{equation}
\label{go:epsilon}
	\epsilon \equiv {E_{tot}/t_{rh}\over E_c/t_{rc}}
\end{equation}
should describe the degree of stability universally (regardless of the presence
of mass spectrum), where
\begin{equation}
\label{go:Ec}
	E_c \equiv {2 \pi\over3} \rho_c r_c^3 v_c^2
\end{equation}
is the energy of the core, and $t_{rh}$ and $t_{rc}$ are half mass and core
relaxation times, respectively.  The motivation for this idea is that the
core luminosity $L_c$ is stabilizing: equations (\ref{rc}) and
(\ref{Lc2}) indicate that $L_c$ will increase as the core shrinks and
decrease as it expands.
The stabilizing influence will be ineffective if $\epsilon\ll 1$, however,
because then the ``equilibrium'' luminosity of the core 
($\propto E_{\rm tot}/t_{rh}$) is very small compared to the
rate at which heat can be removed from the core
if isothermal conditions should break down ($\propto E_{c}/t_{rc}$).

Our two-component models provide a test of these ideas since,
as Goodman (1993) argued, $\epsilon$ depends
both on $N$ and on $m_2/m_1$.
We have performed another set of runs (Group B) whose
parameters were chosen to test the analytical predictions given below.
These simulations include heating by three-body binaries only.
The initial conditions were Plummer models with 3 different mass ratios
($m_2/m_1=$ 1.5, 2, 3) and 4 different total numbers ($N = 3 \times 10^4,
1 \times 10^5, 3 \times 10^5, 1 \times 10^6$).
We have fixed $N_1/N_2 = 100$ except for four supplementary runs.
The parameters of the Group B runs are shown in Table \ref{table:B}.

\begin{deluxetable}{lrrrccc}
\tablecolumns{7}
\tablewidth{0pt}
\tablecaption{Parameters and Results of Simulation Group B \label{table:B}}
\tablehead{
\colhead{Run} &
\colhead{$m_2 \over m_1$} &
\colhead{$N_1 \over N_2$} &
\colhead{$N$} &
\colhead{$\epsilon$} &
\colhead{$r_c \over r_h$} &
\colhead{Oscillation?}
}
\startdata
go01 &1.5&100&$3\times 10^4$& $1.09\times 10^{-2}$ & $4.09\times 10^{-3}$ & n\nl
go02 &2.0&100&$3\times 10^4$& $2.40\times 10^{-2}$ & $7.03\times 10^{-3}$ & n\nl
go03 &3.0&100&$3\times 10^4$& $5.96\times 10^{-2}$ & $1.31\times 10^{-2}$ & n\nl
go04 &1.5&100&$        10^5$& $4.97\times 10^{-3}$ & $1.69\times 10^{-3}$ & y\nl
go05 &2.0&100&$        10^5$& $1.08\times 10^{-2}$ & $2.94\times 10^{-3}$ & n\nl
go06 &3.0&100&$        10^5$& $2.53\times 10^{-2}$ & $5.38\times 10^{-3}$ & n\nl
go07 &1.5&100&$3\times 10^5$& $2.13\times 10^{-3}$ & $6.47\times 10^{-4}$ & y\nl
go08 &2.0&100&$3\times 10^5$& $4.61\times 10^{-3}$ & $1.13\times 10^{-3}$ & y\nl
go09 &3.0&100&$3\times 10^5$& $1.12\times 10^{-2}$ & $2.33\times 10^{-3}$ & n\nl
go10 &1.5&100&$        10^6$& $7.73\times 10^{-4}$ & $2.09\times 10^{-4}$ & y\nl
go11 &2.0&100&$        10^6$& $1.80\times 10^{-3}$ & $4.25\times 10^{-4}$ & y\nl
go12 &3.0&100&$        10^6$& $4.08\times 10^{-3}$ & $8.23\times 10^{-4}$ & y\nl
\tablevspace{0.5ex}
\multicolumn{7}{c}{Supplementary Runs}\nl
\tablevspace{0.5ex}
go05a&2.0& 30&$        10^5$& $0.97\times 10^{-2}$ & $2.70\times 10^{-3}$ & y\nl
go05b&2.0&300&$        10^5$& $0.90\times 10^{-2}$ & $2.57\times 10^{-3}$ & n\nl
go08a&2.0& 30&$3\times 10^5$& $3.98\times 10^{-3}$ & $1.05\times 10^{-3}$ & y\nl
go08b&2.0&300&$3\times 10^5$& $4.15\times 10^{-3}$ & $1.05\times 10^{-3}$ & y\nl
\tablecomments{$\epsilon$ and $r_c/r_h$ values
are equilibrium values.}
\enddata
\end{deluxetable}

\begin{figure}[t]
\plotone{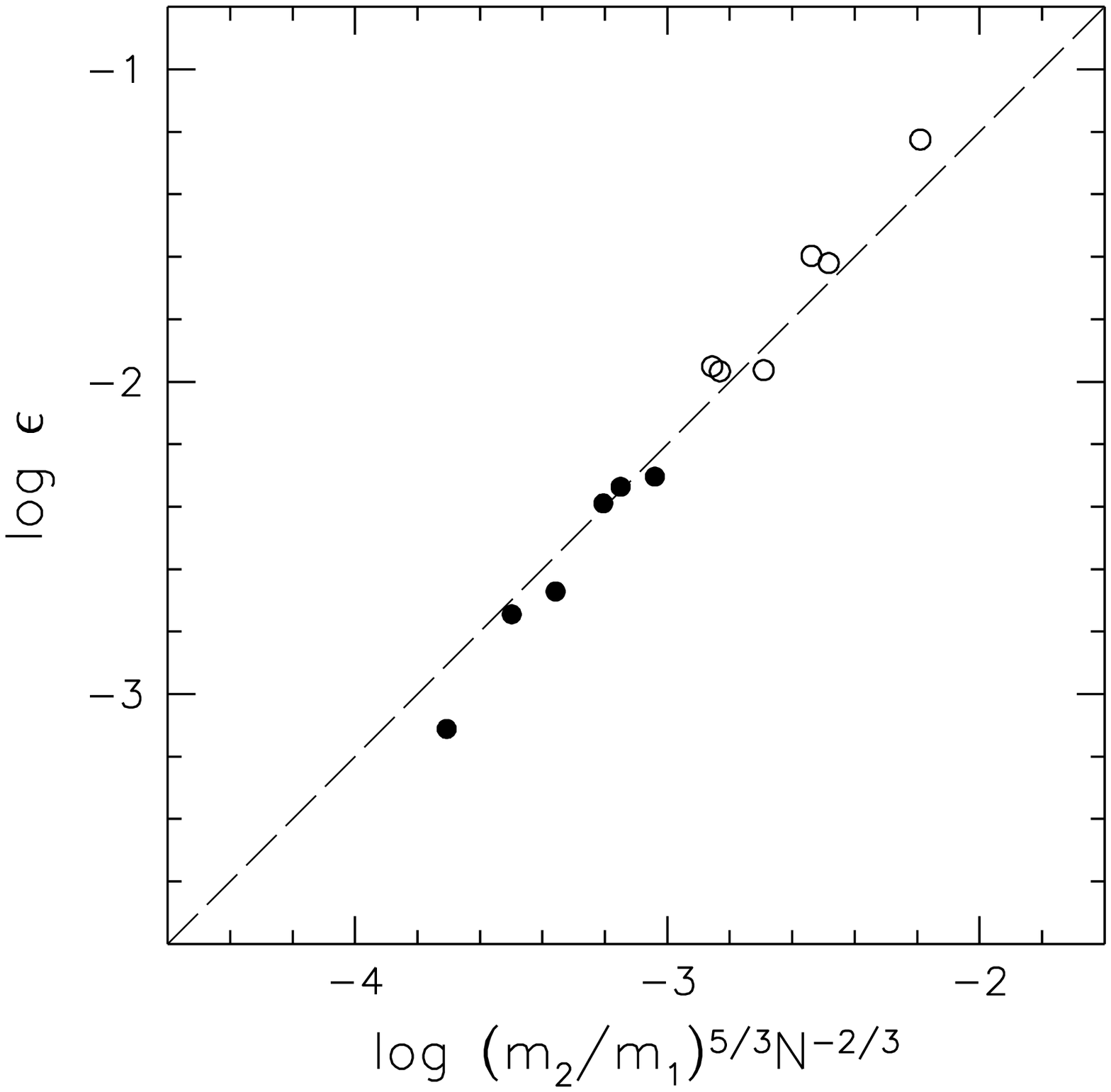}
\caption
{\label{fig:epsil}Oscillation-suppressed $\epsilon$ values of runs
in Group B.  Filled circles are for runs that
showed gravothermal oscillation and open circles are for those that did not.
There is a boundary near $\epsilon \simeq 0.01$ below which oscillations
take place. The diagonal straight line represents the simple scaling
relation in equation (\protect\ref{go:epsil_prop}).
}
\end{figure}

\begin{figure}[t]
\plotone{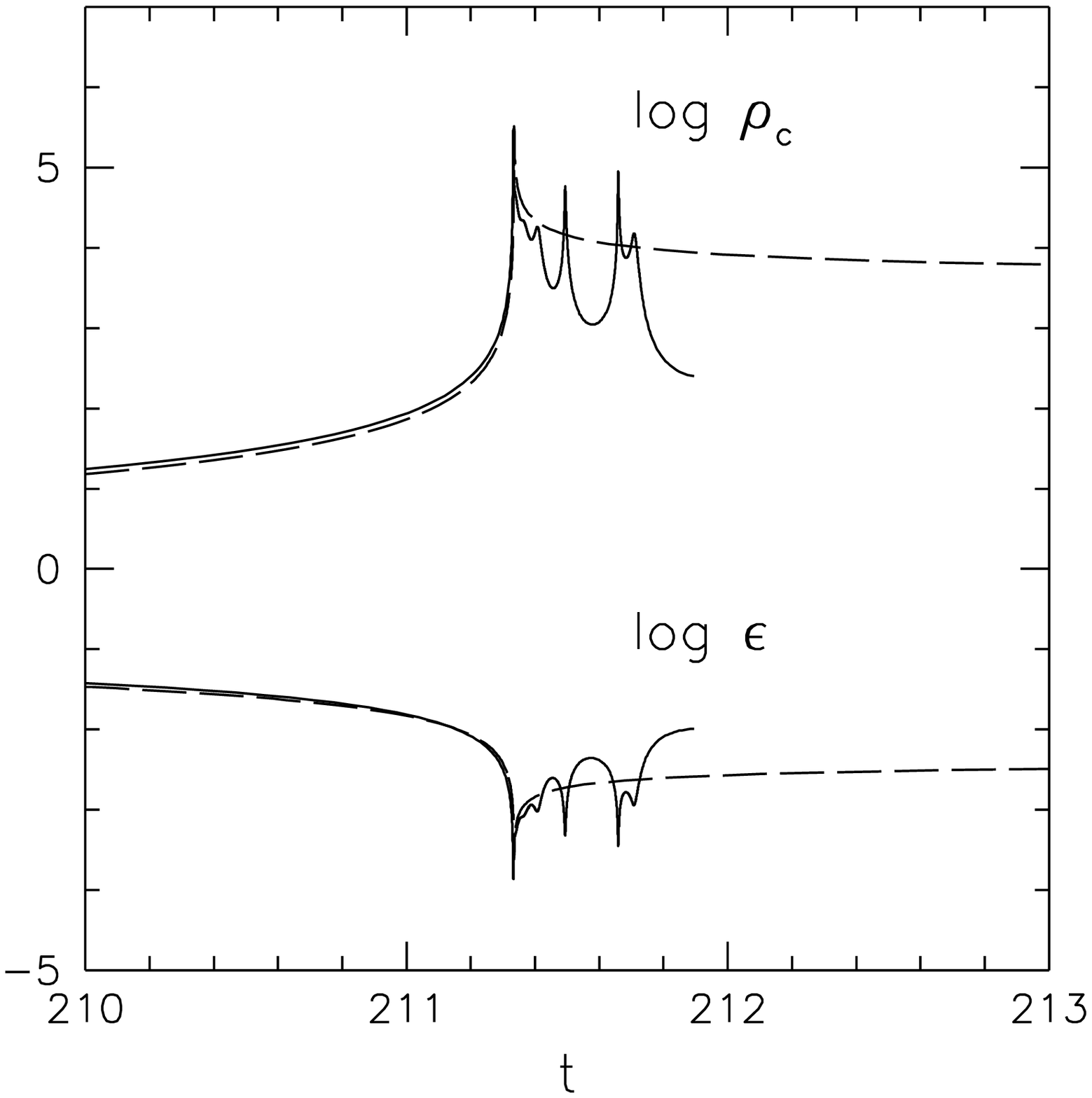}
\caption
{\label{fig:rho+epsilon}Comparison of oscillation-suppressed (dashed
lines) and unsuppressed (solid lines) evolution of central density (upper lines)
and $\epsilon$ (lower lines) of run go12.
The oscillation-suppressed values are obtained by
appropriately enlarging integration timesteps for the Fokker-Planck equation.
The oscillation-suppressed values for $\epsilon$ are good (and almost constant)
representatives of $\epsilon$ during oscillation.
$\rho_c$ is in units of $10^5 M_\odot {\rm pc^{-3}}$ and $t$ in code units
($\simeq t_{rh,i}$).}
\end{figure}

As in Table \ref{table:B} and Figure \ref{fig:epsil}, 6 out of 12 runs
showed gravothermal oscillations in the postcollapse expansion phase.
Figure \ref{fig:rho+epsilon} shows that
$\epsilon$ itself oscillates.
An ``equilibrium'' value for $\epsilon$ can be obtained by adopting
integration time steps larger than the typical oscillation period;
with the implicit time integration schemes that we and others use,
large time steps suppress the gravothermal oscillations.
The equilibrium $\epsilon$ is almost
constant during the whole instability period (Figure
\ref{fig:rho+epsilon}) and can be used as
a representative value for the postcollapse phase.
These representative $\epsilon$'s are plotted in Figure \ref{fig:epsil}
against a combination of cluster parameters that will be explained in the
following subsection.
There is a clear boundary of stability at
$\epsilon_{crit}\approx 0.01$, close to the value $0.013$
found by Goodman (1993) for single-component clusters.

It is instructive to compare certain rows in Table 2.  Runs go04,
go05, and go06 have the same number of stars in the heavier component
but increasing values of $m_2/m_1$, $\epsilon$, and $r_c/r_h$.  The
first run shows oscillations, and the latter two do not.  Therefore,
the suggestion of Murphy, Cohn, \& Hut (1990) that stability depends
on the number of heavy stars is not borne out by these runs.  Note by
the way that the number of heavy stars is only $1000$, much less than
the critical value for instability in single-mass clusters ($7000$).
It is also interesting to compare the unstable run go04 with run go09:
the latter has three times as many stars in each component but is
stable, apparently because of its larger $m_2/m_1$.

\subsection{Dependence of $\epsilon$ on Cluster Parameters}

In their study on the stability of clusters with three-body binaries and
a broad mass spectrum, Murphy et al. (1990) found that stability
persists to much larger $N$ than in single-mass clusters if the mass
function is steep.
They suggested that stability depends mainly on the total number of the
heaviest stars.  We believe that $\epsilon\lesssim 0.01$ is 
a more reliable criterion for the onset of instability.

We now determine $\epsilon$ as a function of cluster parameters.
We derive simple scaling relations by analytic arguments and compare
the results with our numerical results.

First we reproduce the prediction by Goodman (1993).
From equation (\ref{scale}), one obtains
\begin{equation}
\label{go:r_prop}
	\left ( {r_c \over r_h} \right)  \propto  \left ( {m_2 \over m_1}
		\right )^{7/6} N^{-2/3}.
\end{equation}
Now, from equations (\ref{Mc}), (\ref{rc}), and (\ref{vc}), the ratio
of energies in the cluster and the core is
\begin{equation}
\label{go:Eratio}
	{E_{tot} \over E_c} \propto {M \over M_c} {v_m^2 \over v_c^2}
			    \propto {r_h \over r_c} {v_m^4 \over v_c^4},
\end{equation}
where $v_m^2$ is the velocity dispersion of the whole cluster.  Similarly,
the ratio of half-mass to core relaxation time is
\begin{equation}
\label{go:tratio}
	{t_{rh} \over t_{rc}} \propto {M^{1/2} \over M_c^{1/2}}
			       {r_h^{3/2} \over r_c^{3/2}}
			       {\bar m_c \over \bar m }
			      \propto {r_h^2 \over r_c^2} {v_m \over v_c}
			       {\bar m_c \over \bar m},
\end{equation}
where $\bar m$ is the mean mass in the cluster, and $\bar m_c$ is the
mean mass in the core.  With $\bar m_c / \bar m \approx m_2/m_1$
and $v_m^2/v_c^2 \approx m_2/m_1$, one finally obtains
\begin{eqnarray}
\label{go:epsil_prop}
	\epsilon & \propto & \left ( {r_c \over r_h } \right )
			   \left ( {m_2 \over m_1} \right )^{1/2} \nonumber\\
		 & \propto & \left ( {m_2 \over m_1} \right )^{5/3} N^{-2/3}.
\end{eqnarray}
The exponents of mass ratio $3/2$ and 2 in equation (18) of Goodman (1993)
should be corrected to $7/6$ and $5/3$ as in equations (\ref{go:r_prop}) and
(\ref{go:epsil_prop}).

We have compared the above scaling relations with our numerical results from
runs in Group B.  The equilibrium $\epsilon$ values of our runs are plotted
in Figure \ref{fig:epsil} against the righthand side of equation
(\ref{go:epsil_prop}).  The data points are well aligned and their slope
is $1.20 \pm 0.05$, slightly higher than the theoretical value
of unity.  

\begin{figure}[t]
\plotone{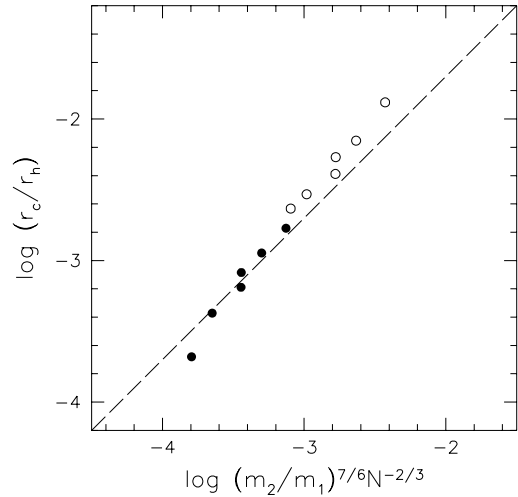}
\caption
{\label{fig:rcrh}Oscillation-suppressed $r_c/r_h$ values of our runs
in Group B.  Filled circles are for runs that
showed gravothermal oscillation and open circles are for those that did not.
The straight diagonal line represents the theoretical scaling
relation.}
\end{figure}

This discrepancy is traceable to the deviation of
$r_c/r_h$ from the predicted scaling (\ref{go:r_prop}).
In \S \ref{sec:evolution}, we remarked that our numerical results show a
slightly higher slope than expected for $r_c$ ($1.07 \pm 0.01$ versus $1$)
and a lower slope for $r_h$ ($0.86 \pm 0.06$ versus $1$).
Thus the slope for $r_c/r_h$ should be higher than predicted by our
analytic arguments by appproximately $0.21$ for the runs in Group A.
The correlation between the two sides
of equation (\ref{go:r_prop}) for Group B is shown in Figure \ref{fig:rcrh}.
The slope of these data points is $1.26 \pm 0.04$, which is comparable to
the one for $\epsilon$.
We find that one of the causes of this poor $r_c/r_h$ approximation is
the assumption that the proportionality constant in equation (\ref{rh-t}) is
independent of cluster parameters.  However, we find that the ratio
$(r_h/t_{rh})/ \dot r_h$ of our runs ranges from 10 to 15.  Runs with
larger $m_2/m_1$ show larger $(r_h/t_{rh})/ \dot r_h$ values and this
$m_2/m_1$ dependence is more pronounced for runs with larger $N$.  If one lets
\begin{equation}
\label{coef_A}
	{r_h \over t_{rh}} \equiv A \, \dot r_h,
\end{equation}
where the coefficient $A$ depends on cluster parameters such as
$m_2/m_1$ and $N$, one finds
\begin{equation}
\label{coef_A2}
	r_c/r_h \propto A^{8/9}.
\end{equation}
Then a 23 \% variation in $A$ would give a 20 \% residual in the
slope of the correlation between the two sides of equation (\ref{rh-t}).
This is approximately what one sees in Figure (8), which is based on
$m_2/m_1$ ranging from $1.5$ to $3$.  Although equation (\ref{rh-t}) 
has been used widely without consideration of its dependence on
cluster parameters, we conclude that the coefficient in this
relation varies with the mass function.

We have made four more runs in addition to Group B to test the dependence of
$\epsilon$ on $N_1/N_2$, which is not apparent in equation
(\ref{go:epsil_prop}): two runs with the same parameters as run go05 except
$N_1/N_2$ = 30 and 300 (go05a and go05b, respectively), 
and two runs as run go08
except $N_1/N_2$ = 30 and 300 (go08a and go08b, respectively).
Equilibrium $\epsilon$ values of these runs are $0.97 \times 10^{-2}$ for
run go05a, $0.90 \times 10^{-2}$ for run go05b, $3.98 \times 10^{-3}$ for
run go08a, and $4.15 \times 10^{-3}$ for run go08b.  These values are all
within only 15 \% differences from their comparison runs, go05 and go08,
indicating that $\epsilon$ is independent of $N_1/N_2$ as expected in
the above energy balance analysis.  However, while gravothermal oscillations
are observed in run go08a and not in runs go05a and go05b as expected,
it is not observed in run go08b.  Clearly the criterion of 
$\epsilon \ga 0.01$
does not appear to be exact if $N_2$ is too small. 

All of our previous analyses assume that there are enough stars in the
heavy component so that the core is dominated by them. 
This requires that the ratio of core mass to total mass,
\begin{equation}
M_c/M \approx 3.3\times \left(
{m_2\over m_1}\right)^{1/6} N^{-2/3} t_{11}^{-0.05}
\end{equation}
(see eq.~[\ref{eq:mcore}])
should be smaller than $M_2/M$. In order to confine
most of the heavy stars to a region much smaller than $r_h$, we also 
require $m_2/m_1>3/2$: $\rho_2(r)\propto \rho_1(r)^{m_2/m_1}$
in equipartion; $\rho_1\propto r^{-2}$ in the regions well outside
the core where the lighter component dominates the potential; and we
require $\rho_2(r)$ to fall more steeply than $r^{-3}$ in those regions
so that most of the heavies are at smaller radii.  All of our runs 
satisfy these requirements, except where $m_2/m_1=1.5$ so that the
second condition is marginally violated.

\section{SUMMARY}
\label{sec:summary}

We have investigated the evolution of isolated two-component
clusters with Plummer-model initial conditions. We have
included the main effects of  both three-body and tidal-capture binaries
by adding a heating term to the Fokker-Planck equation. 

In agreement with previous investigators, we find that core collapse
is hastened by the presence of heavy remnant stars. 

In the postcollapse phase, we find that heating by three-body binaries
exceeds that by tidal binaries at least for clusters with
$M \le 3 \times 10^5 \msun$, $r_{h,i} \ge 5 \, {\rm pc}$,
$m_2/m_1 \ge 2$, and $N_1/N_2 \le 300$ when $m_2=1.4\msun$.
When three-body binary heating does dominate,
the expansion of the postcollapse cluster is self-similar. 
Scaling laws for cluster parameters including the
central density, velocity dispersion, core radius, and half-mass radius
have been derived from simple considerations of energy balance, and
these scalings generally agree well with our numerical results, which
however also provide the numerical coefficients in the scaling laws.
We related the postcollapse evolution of these cluster parameters
to $N$, $M$ and $m_1/m_2$. 

We have studied the gravothermal oscillation phenomenon using our two-component
models. We have confirmed 
that the parameter $\epsilon$ = $(E_{tot}/t_{rh})/
(E_c/t_{rh})$ predicts the occurence of gravothermal oscillations in
the presence of this simplest of nontrivival mass functions.
The scaling law for $\epsilon$ with respect to $m_1/m_2$ and $N$ is derived
in the limit of small $N_2/N_1$ and compared 
with our numerical results. 
Generally speaking, clusters with a 
steeper mass function are less susceptible to gravothermal instability.
This is in qualitative agreement with an earlier suggestion by
Murphy et al. (1990), but whereas these authors suggested
that stability depends on the number of heavy stars, we conclude that
it is independent of the number of heavies but depends jointly
on the total number of stars and on the ratio of the individual
stellar masses in the two components.

These conclusions need to be tested against models with
more mass components. Our preliminary studies indicate
that the evolution of multi-mass clusters is generally similar to 
two-component models for an appropriate choice of model parameters 
(mostly suitable $m_2/m_1$). 
In order to compare with real clusters, we need to take into account
a host of complicating physical effects, most importantly an
external tidal field. The results of these studies will be reported
elsewhere.

\acknowledgements
S. S. K. thanks Chang Won Lee and Jung-Sook Park for obtaining old and rare
papers.  This research was supported in part by Korea Science and Engineering
Foundation, 95-0702-01-01-03, and in part by the Matching Fund Programs of
Research Institute for Basic Sciences, Pusan National University,
RIBS-PNU-96-501.



\begin{references}
\reference{A62} Antonov, V. A. 1962, {\it Vest. Leningrad Univ.}, 7, 135
(English transl. in IAU Symposium 113, Dynamics of Star Clusters, ed.
J. Goodman and P. Hut [Dordrecht: Reidel], p. 525 [1985])
\reference{BH87} Benz, W. \& Hills, J. G. 1987, ApJ, 323, 614
\reference{BS84} Bettwieser, E., \& Sugimoto, D. 1984, \mnras, 208, 493
\reference{CW90} Chernoff, D. F. \& Weinberg, M. 1990, \apj, 351, 121
\reference{C79} Cohn, H. N. 1979, ApJ, 234, 1036
\reference{C80} Cohn, H. N. 1980, ApJ, 242, 765
\reference{C85} Cohn, H. N. 1985, in IAU Symposium 113, Dynamics of star
clusters, ed. J. Goodman and P. Hut (Dordrecht: Reidel), p. 161
\reference{G95} Drukier, G. A., 1995, ApJS, 100, 347
\reference{GGCM91} Gao, B., Goodman, J., Cohn, H., \& Murphy, B. 1991, ApJ,
370, 567
\reference{GF95} Gebhardt, K., \& Fischer, P. 1995, AJ, 109, 209
\reference{GO97} Gnedin, O. Y. \& Ostriker, J. P. 1997, ApJ, 474, 223
\reference{G87} Goodman, J. 1987, ApJ, 313, 529.
\reference{G93} Goodman, J. 1993, in Structure and Dynamics of Globular
Clusters, ASP Conference Series Vol. 50, eds. S. G. Djorgovski, \& G. Meylan
(San Francisco: ASP), p. 87
\reference{GYSB96} Guhathakurta, P., Yanny, B., Schneider, D. P., \& Bahcall,
J. N., 1996, AJ, 111, 267.
\reference{H61} H\'enon, M., 1961, {Ann. d'Ap.}, 24, 369
\reference{I85} Inagaki, S. 1985, in IAU Symposium 113, Dynamics of star
clusters, ed. J. Goodman and P. Hut (Dordrecht: Reidel), p. 189
\reference{IW84} Inagaki, S., \& Wiyanto, P. 1984, PASJ, 36, 391
\reference{KL97} Kim, S. S., \& Lee, H. M. 1997, J. of Korean Astron. Soc.,
vol. 30, no. 2, in press
\reference{KHM} Kulkarni, S. R., Hut, P., \& MaMillan, S. 1993, Nature, 364, 421
\reference{KO95} Kundic, T. \& Ostriker, J. P. 1995, ApJ, 438, 702
\reference{L87} Lee, H. M. 1987, ApJ, 319, 772
\reference{LFR91} Lee, H. M., Fahlman, G. G., \& Richer, H. B. 1991, ApJ, 366,
455
\reference{LG95} Lee, H. M., \& Goodman, J. 1995, ApJ, 443, 109
\reference{LO86} Lee, H. M., \& Ostriker, J. P. 1986, ApJ, 310, 176
\reference{LO93} Lee, H. M., \& Ostriker, J. P. 1993, ApJ, 409, 617
\reference{LW68} Lynden-Bell, D., \& Wood, R. 1968, MNRAS, 138, 495
\reference{MCH90} Murphy, B. W., Cohn, H. N., \& Hut, P. 1990, MNRAS, 245, 335
\reference{P93} Phinney, E. S. 1993, in Structure and Dynamics of Globular
clusters, S. G. Djorgovski \& G. Meylan, eds. (San Francisco, ASP), p. 137
\reference{Q96} Quinlan, G. D. 1996, New Astronomy, 1, 255
\reference{S87} Spitzer, L. Jr. 1987, Dynamical Evolution of Globular
Clusters (Princeton: Princeton University Press)
\reference{SOC87} Statler, T. S., Ostriker, J. P., \& Cohn, H. N. 1987, ApJ,
316, 626
\reference{SB83} Sugimoto, D., \& Bettwieser, E. 1983, MNRAS, 204, 19P
\reference{TLI95} Takahashi, K., Lee, H. M., \& Inagaki, S. 1997, MNRAS,
submitted
\reference{Y78} Yoshizawa, M., Inagaki, S., Nishida, M. T., Kato, S., Tanaka,
Y., \& Watanabe, Y. 1978, PASJ, 30, 279
\end{references}
\end{document}